\documentclass{bmvc2k}

\title{SR+Codec: a Benchmark of Super-Resolution for Video Compression Bitrate Reduction}

\addauthor{Evgeney Bogatyrev}{evgeney.zimin@graphics.cs.msu.ru}{1}
\addauthor{Ivan Molodetskikh}{ivan.molodetskikh@graphics.cs.msu.ru}{1}
\addauthor{Dmitriy Vatolin}{dmitriy@graphics.cs.msu.ru}{1,2}

\addinstitution{
 Lomonosov Moscow State University,\\
 Moscow, Russia
}
\addinstitution{
 MSU Institute for Artificial Intelligence,\\
Moscow, Russia
}

\runninghead{Bogatyrev \etal}{SR for Video Compression Benchmark}

\def\etal{\emph{et al}\bmvaOneDot}

\usepackage{svg}
\usepackage{slashbox}
\usepackage{capt-of}
\usepackage{comment}
\usepackage{graphicx}
\usepackage{pdfpages}

\begin{document}

\maketitle

\begin{abstract}
In recent years, there has been significant interest in Super-Resolution (SR), which focuses on generating a high-resolution image from a low-resolution input. Deep learning-based methods for super-resolution have been particularly popular and have shown impressive results on various benchmarks. However, research indicates that these methods may not perform as well on strongly compressed videos.

We developed a super-resolution benchmark to analyze SR's capacity to upscale compressed videos. Our dataset  employed video codecs based on five widely-used compression standards: H.264, H.265, H.266, AV1, and AVS3. We assessed 19 popular SR models using our benchmark and evaluated their ability to restore details and their susceptibility to compression artifacts. To get an accurate perceptual ranking of SR models, we conducted a crowd-sourced side-by-side comparison of their outputs. We found that some SR models, combined with compression, allow us to reduce the video bitrate without significant loss of quality. We also compared a range of image and video quality metrics with subjective scores to evaluate their accuracy on super-resolved compressed videos. The benchmark is publicly available at \href{https://videoprocessing.ai/benchmarks/super-resolution-for-video-compression.html}{videoprocessing.ai/benchmarks/super-resolution-for-video-compression.html}.
\end{abstract}
\section{Introduction}
\label{sec:intro}
Super-resolution (SR) is the task of increasing the resolution of images and videos, with potential use ranging from detail restoration to quality enhancement~\cite{liang2022recurrent,wang2021realesrgan}. Some state-of-the-art SR methods can restore details not clearly visible in the original (lower-resolution) clip while working in real time~\cite{wu2022animesr}. Neighboring frames can help fill gaps when upscaling, because small movements caused by camera tremor may provide enough information to accurately increase the resolution, as demonstrated using a Google Pixel 3 camera \cite{10.1145/3306346.3323024}. This ability of SR methods to restore video details and enhance video quality suggests their use for improving video compression efficiency.

Video traffic had accounted for more than 80\% of all consumer Internet traffic by 2022~\cite{cisco} and that number is continuing to rise. Video compression that can lower bandwidth consumption with minimal changes to the visual quality of the video is more critical than ever. There have been some breakthroughs in video compression techniques, such as the recently developed MPEG compression standard MPEG-5 Part 2 Low Complexity Enhancement Video Coding (LCEVC)~\cite{meardi2020mpeg}. The core idea of LCEVC is to use a conventional video codec at a lower resolution and reconstruct a full-resolution video after decoding using enhancement layers, leading to a significant decrease to bandwidth consumption.

\begin{figure}[t]
\centering
\includegraphics[width=1.0\linewidth]{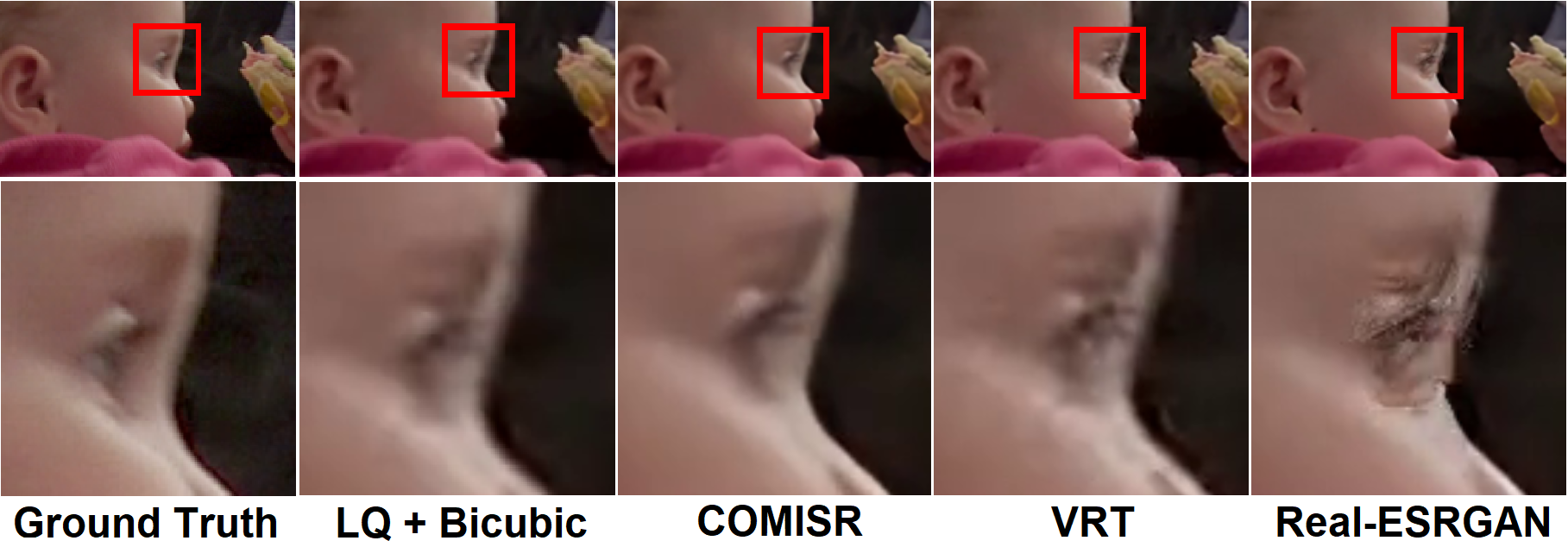}
\caption{The comparison between 3 super-resolution models on compressed video sequence. COMISR eliminates the compression artifact because it is designed to work with compressed video. On the other hand, VRT and Real-ESRGAN fail to remove this artifact.}
\label{fig:preview}
\end{figure}

Some recent codecs~\cite{meardi2020mpeg, khani2021efficient} downscale a video before compression to cut the bitrate and then upscale it to its original resolution using SR. Not all SR methods are suitable for such downscale-based video compression, however, since few real-time SR models can generate acceptable-quality video. Our research shows that many SR models are unable to deal with compression artifacts, as shown in Figure \ref{fig:preview}. Several existing benchmarks assess SR methods' ability to upscale compressed videos~\cite{yang2022ntire,yang2022aim}, focusing on perceptual quality. We improve upon this work by additionally considering video bitrate reduction.

To analyze which SR models work better with each compression standard, and to help researchers find the best models for their codecs, we present our Super-Resolution for Video Compression benchmark. To develop it we selected 19 popular SR models with different architectures and assessed their compressed-video-restoration capabilities on our dataset, which includes videos compressed using five codecs. Our effort employed objective metrics and subjective evaluation to assess model quality. In addition, we analysed the correlation between objective-quality metrics and subjective scores and calculated bitrate reduction that can be achieved using each SR model during the compression.

Our main contributions are as follows:\begin{enumerate}
    \item We present a comprehensive SR benchmark to test the ability of 19 SR models to upscale and restore videos compressed by five video codecs of different standards. We evaluate the perceptual quality of the restored videos by conducting a crowd-sourced subjective comparison with 5397 subjects. The benchmark is publicly available at \href{https://videoprocessing.ai/benchmarks/super-resolution-for-video-compression.html}{https://videoprocessing.ai/benchmarks/super-resolution-for-video-compression.html}.
    \item For every tested codec, we show which SR methods provide the most video bitrate reduction with the least quality loss. We find bitrate improvements of up to 65\% compared to using the base codec directly without SR.
    \item We analyse six video quality metrics by their correlation with subjective scores on our dataset. Based on their performance on individual video clips, we construct a simple metric combination with improved results.
\end{enumerate}
\section{Related Work}
\label{sec:relatedwork}
In this section we provide an overview of existing SR methods, downscale-based video codecs, and SR benchmarks.

\subsection{Super-resolution methods}
SR has received extensive attention since neural networks were first applied to this area, resulting in many approaches.  

Some SR algorithms rely on the temporal redundancy of video frames, allowing them to restore a single high-resolution frame from a series of low-resolution ones. \textbf{RBPN}~\cite{haris2019recurrent} integrates spatial and temporal contexts from a video using a recurrent encoder-decoder module. \textbf{COMISR}~\cite{li2021comisr} upscales compressed videos; it employs bidirectional recurrent warping for detail-preserving flow estimation, and it applies Laplacian enhancement. \textbf{BasicVSR++}~\cite{chan2022basicvsr++} also adopts bidirectional propagation and spatial alignment. \textbf{VRT}~\cite{liang2022vrt} extracts video features, upscales them, and then reconstructs HQ frames on the basis of these features using a transformer network. \textbf{RVRT}~\cite{liang2022recurrent} divides videos into multiple clips and uses previously inferred clip features to estimate subsequent clip features. Additionally, a guided deformable attention mechanism facilitates clip-to-clip alignment by predicting and aggregating multiple relevant locations across different clips. \textbf{Swin2SR}~\cite{conde2022swin2sr} explores Swin Transformer V2 to improve \textbf{SwinIR}~\cite{liang2021swinir} compressed image super-resolution. 

Generative adversarial networks (GANs) serve widely in deep learning and especially in SR. \textbf{ESRGAN}~\cite{wang2018esrgan} modifies the SRGAN architecture by adding residual-in-residual dense block as well as improved adversarial loss and perceptual loss. \textbf{Real-ESRGAN}~\cite{wang2021realesrgan} enhances this approach by incorporating high-order degradation modeling to simulate real-world degradation.

Lately, diffusion models have demonstrated impressive abilities to generate high-quality results in various applications, and SR is no exception~\cite{saharia2022image,yue2024resshift,wang2023sinsr,noroozi2024you}. Although diffusion-based models have shown promising results, their drawback lies in the long inference process. These models usually need multiple inference steps to produce a final output, which hinders their practical use. Since they are unable to work in real time, they can't be used in the video decoding process. Therefore, in this study we are not considering diffusion models.

Given the limited number of SR models designed to work with compressed video, assessing the performance of existing SR models on compressed video remains a critical task.

\subsection{Downscale-based video codecs}
Recently, some video codecs have been designed to downscale the video before compression to reduce the bitrate, and then upscale it to the original resolution on the decoder side. Researchers are exploring many approaches to the upscaling module, ranging from simple filters to extensive neural networks. 

The core idea of \textbf{LCEVC}~\cite{meardi2020mpeg} is to use a conventional video codec at a lower resolution and reconstruct a full-resolution video by combining the decoded low-resolution video with up to two enhancement sub-layers of residuals encoded with specialized low-complexity coding tools. Some video codecs use the similar idea of implementing a super-resolution network on the decoder side.

\textbf{SRVC}~\cite{khani2021efficient} encodes video into two bitstreams: a content stream and a model stream. The content stream is produced by compressing downsampled low-resolution video with the existing codec. The model stream encodes updates to the lightweight super-resolution network, which is used to upscale video at the decoder side. The SR network is trained on local segments of the video during the encoding process.

\textbf{RR-DnCNN}~\cite{ho2021rr} addresses the problem of removing compression artifacts during the downscale-based video coding. A straightforward approach of applying compression artifact removal techniques before SR may result in detail loss. The paper proposes an end-to-end restoration-reconstruction deep neural network using the degradation-aware technique.

\textbf{ViSRTA}~\cite{afonso2018video} takes one step further and dynamically resamples the video not only spatially, but also temporally
during the encoding process. CNN-based architecture is used to restore spatial resolution, and temporal upsampling is performed by frame repetition.

\textbf{Lin \etal}~\cite{lin2021cnn} proposes a CNN-based SR method for resample-based video coding on the basis of the VTM codec. The authors designed separate networks for the luma and chroma components which exploit the cross-component correlation by using the luma reconstruction as the auxiliary information for the chroma network.

In our benchmark we take the idea of these codecs and try to apply it to different combinations of widely-used codecs and SR methods. Downscale-based video codecs are usually designed to transmit extra features of the original video alongside the low-resolution video data, to aid in the upscaling process during decoding. In our benchmark, codecs and SR methods work independently of each other, as our main goal is to evaluate how different codec and SR methods fit together.

\subsection{Super-resolution benchmarks}
Many SR benchmarks and challenges have appeared recently. We focus our attention on the ones prioritizing super-resolving compressed videos.

\textbf{NTIRE 2022} Challenge on Super-Resolution and Quality Enhancement of Compressed Video~\cite{yang2022ntire} presents evaluation results of quality enhancement of compressed videos (track 1), quality enhancement with 2× and 4× SR (track 2 and track 3 respectively). More than 600 participants have registered on all tracks in total. This challenge proposes LDV 2.0 dataset with 335 videos and uses PSNR to evaluate participants. 

The \textbf{AIM 2022} Challenge on Super-Resolution of Compressed Image and Video~\cite{yang2022aim} presents two tracks: restoration of compressed images and compressed video. PSNR is used to evaluate the results, and an extension of the previously mentioned LDV 2.0 dataset, the LDV 3.0 dataset of 365 videos, is proposed. The second track of this task requires participants to enhance and 4$\times$ super-resolve HEVC compressed videos.

Our benchmark improves upon all the benchmarks presented above by a wider set of video quality metrics and a variety of video codecs used to create distorted videos. We also measure the bitrate reduction that SR models can provide when used during decompression.
\section{Benchmark Methodology}

In this section, we describe the process of preparing the dataset for our benchmark, including the methodology for selecting videos and preprocessing them. We also describe our benchmark evaluation procedure and quality assessment.


\begin{figure}
\begin{minipage}[c]{0.60\linewidth}
\vspace{0pt}
  \centering
  \includegraphics[width=.9\linewidth]{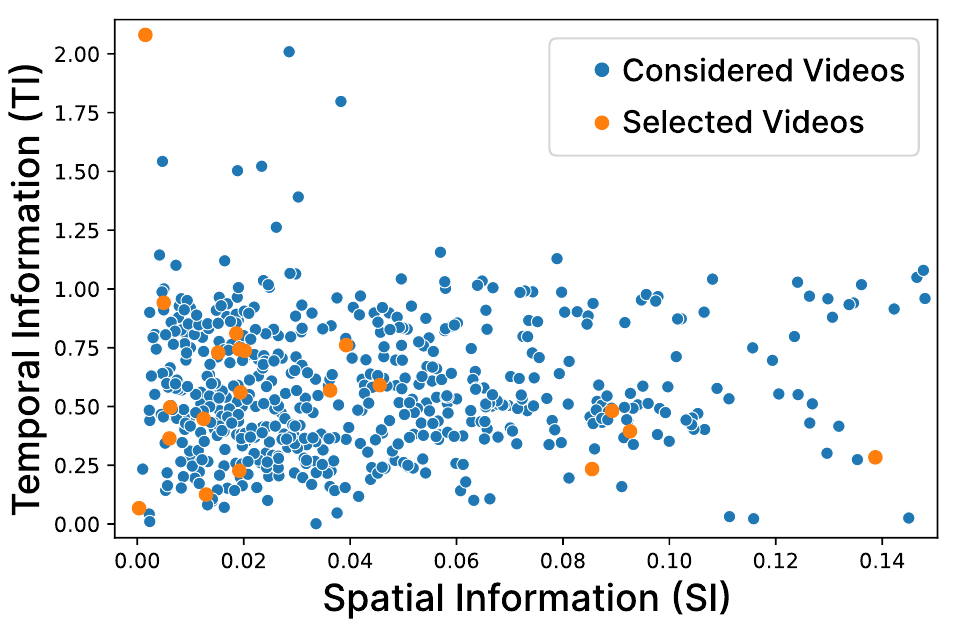}
  \caption{Distribution of Google Spatial and Temporal~\cite{wang2019youtube} information for videos we considered when creating our training dataset. Chosen videos appear in orange, others in blue.}
  \label{fig:si_ti_dist}
\end{minipage}%
\hspace{0.025\linewidth}
\begin{minipage}[c]{0.35\linewidth}
\vspace{0pt}
  \centering
  \begin{tabular}{c|c}    
\hline
Codec & Standard \\
\hline
x264 & H.264 \\
x265 & H.265 \\
aomenc & AV1 \\
vvenc~\cite{VVenC} & H.266 \\
uavs3e~\cite{uavs3e} & AVS3 \\
\hline
\end{tabular}
\vspace{5pt}
\captionof{table}{Video codecs used to compress videos from benchmark dataset. We used \textit{medium} preset for all of these codecs.}
\label{tab:codecs_overview}
\end{minipage}
\end{figure}
  
\subsection{Dataset preparation} \label{dataset}
\begin{figure}
\centering
\includegraphics[width=1.0\linewidth]{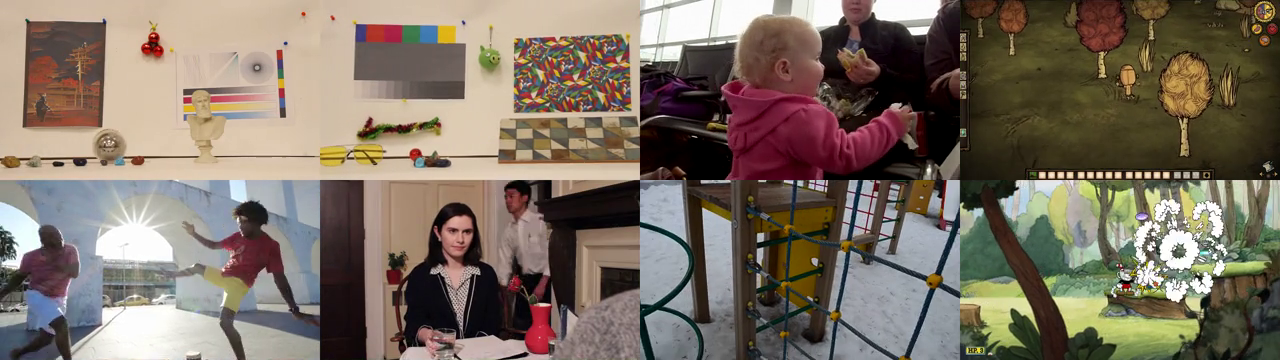}
\caption{Example videos from the dataset. The dataset includes real-world sequences, animation, and clips from games.}
\label{fig:dataset}
\end{figure}

To ensure the benchmark dataset is diverse enough to test various aspects of SR models, we collected $1920\times1080$ videos from multiple sources:
\begin{itemize}
    \item \textbf{Vimeo}: We gathered 50 sequences, including both real world and animation. We split them into scenes using the Scene Change Detector plugin for VQMT~\cite{scd}.
    \item \textbf{Camera}: We shot several videos using a Canon EOS 7D. The settings aimed to minimize blur and achieve the appropriate brightness — the ISO was 4000 and the aperture 400. Those settings provided clear ground-truth (GT) videos without blur or noise. We shot 20 indoor videos and 30 outdoor videos. The indoor ones consist of synthetically crafted scenes containing objects from everyday life. Each scene includes either moving objects or horizontal camera motion.
    \item \textbf{Games}: We recorded 20 clips from various 2D and 3D videogames.
\end{itemize}
We then obtained the following features for each video: Google Spatial and Temporal features~\cite{wang2019youtube}, frames per second (FPS), colorfulness~\cite{hasler2003measuring}, and maximal number of faces~\cite{facerecognition} throughout the video. On the basis of these features we separated all videos into 20 clusters using the K-Means clustering and selected one video from each cluster, as shown in Figure~\ref{fig:si_ti_dist}. We refer to these selections as \textit{source videos}. A preview of them appears in Figure~\ref{fig:dataset}.  

To ensure that important fine details of the video are not completely lost after downscaling and compression, we considered only videos with low space and time complexity, and no major blur or noise. However, camera motion was required, as it can help video SR models restore each frame more faithfully using neighboring frames.

\subsection{Benchmark pipeline}

To select SR models for evaluation we used SR benchmarks that target two tasks: detail restoration~\cite{detailrest} and perceptual-quality improvement~\cite{videoupscalers}. We excluded similar SR methods based on their performance in these benchmarks.

We selected the following 19 methods: BasicVSR++~\cite{chan2022basicvsr++}, COMISR~\cite{li2021comisr}, DBVSR~\cite{pan2021deep}, EGVSR~\cite{cao2021real}, LGFN~\cite{su2020local}, RBPN~\cite{haris2019recurrent}, Real-ESRGAN~\cite{wang2021realesrgan}, RealSR~\cite{ji2020real}, RSDN~\cite{isobe2020video}, SOF-VSR-BD~\cite{wang2018learning}, SOF-VSR-BI~\cite{wang2018learning}, SwinIR~\cite{liang2021swinir}, TMNet~\cite{xu2021temporal}, VRT~\cite{liang2022vrt}, RVRT~\cite{liang2022recurrent}, IART\cite{iart2024}, AnimeSR\cite{wu2022animesr}, Topaz 
Video AI~\cite{topaz}, and bicubic interpolation. For all models we used pretrained weights provided by authors.
Given the scarcity of high-quality video SR models, we decided to include image SR in our comparison.

We employed the standard video codecs for five widely-used compression standards, as detailed in Table \ref{tab:codecs_overview}.

\begin{figure}
\centering
\includegraphics[width=1.0\linewidth]{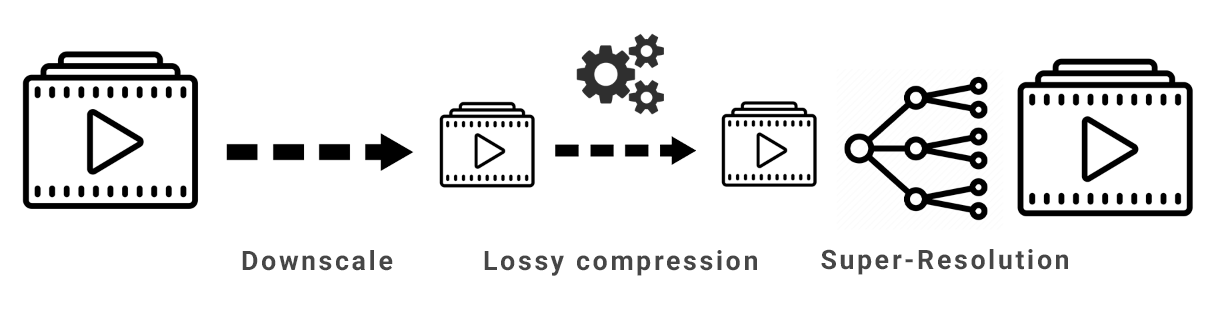}
\caption{The evaluation pipeline of our benchmark. The pipeline consists of three steps: $4\times$ bicubic downscaling, compression, and $4\times$ SR upscaling.}
\label{fig:pipeline}
\end{figure}

A brief visualization of the benchmark pipeline can be seen in Figure \ref{fig:pipeline}. First, we downscaled the source video to $480\times270$ resolution using FFmpeg with the \texttt{flags=bicubic} option. We then compressed the low-resolution video using each of the five video codecs at three target bitrates: 0.6, 1.0, and 2.0 Mbps. To get more accurate objective metric scores, we also compressed videos at 0.1, 0.3, and 4.0 Mbps; however, these bitrates were not used in the subjective evaluation. We chose these bitrates to be relatively low and to form a logarithmic curve. All codecs employed the \textit{medium} preset during compression. Compressed videos underwent transcoding to PNG sequences using FFmpeg, which were then passed as inputs to an SR model. We applied image SR models to each frame individually; video SR models received the path to the directory containing frames in the correct order. We tested a 4× upscale using our benchmark, but some SR models can only handle 2×. In this latter case, we applied the model twice. 

\subsection{Quality estimation and subjective study}
After super-resolving videos, we calculated the following objective-video-quality metrics on the results: PSNR, MS-SSIM~\cite{wang2004image}, VMAF~\cite{VMAF_repo}, LPIPS~\cite{zhang2018unreasonable}, MDTVSFA~\cite{li2021unified} and ERQA~\cite{kirillova2021erqa}. We considered mainly full-reference metrics since we prioritized detail restoration over perceptual quality. PSNR, MS-SSIM, and VMAF are the standard for compressed video quality evaluation~\cite{MSUcodecs}, while LPIPS and ERQA are well-suited for super-resolved images~\cite{detailrest}. The only no-reference metric we used is MDTVSFA, since it shows promising results when evaluating compressed sequences~\cite{NEURIPS2022_59ac9f01}. To rank bitrate reduction, we calculated BSQ-rate (bitrate-for-the-same-quality rate)~\cite{zvezdakova2020bsq} for each SR+codec pair relative to base codec performance, where the base codec is the one we used to compress low-resolution video. A lower BSQ-rate means more bitrate saving for the same quality.

To subjectively rank SR models, we conducted a crowd-sourced comparison through Subjectify.us~\cite{subjectify} service. Because detail loss and compression artifacts can be difficult to notice in a full frame, the subjective evaluation employed crops. We needed pairs of crops to fit on the assessors' screens during the comparison, so we choose the resolution of the crops to be $480\times270$. First, we generated saliency maps for each source video using a method proposed in Kroner \etal~\cite{KRONER2020261}. Second, we averaged the saliency maps over all frames and applied a Gaussian-blur kernel to the result in order to determine the video’s most salient region. Third, we took distorted videos from the benchmark dataset and cut one $480\times270$ crop from each one, with the most salient area at the center of the crop. We evaluated objective metrics on these crops to determine the correlation with the subjective scores.

\begin{table}
\begin{center}
 \resizebox{\textwidth}{!}{\begin{tabular}{ll|ccccc}
\hline
 Codec & SR & Subj. score $\uparrow$  & ERQA $\uparrow$  & LPIPS $\downarrow$  & PSNR $\uparrow$   & MDTVSFA $\uparrow$  \\
\hline
x264  &  SwinIR~\cite{liang2021swinir} & \textbf{3.695} & 0.698 & \textbf{0.206} & 24.895 & \textbf{0.564} \\ 
x264  &  Real-ESRGAN~\cite{wang2021realesrgan} & 3.092 & 0.740 & 0.267 & 26.983 &  0.501 \\
x264  &  \textit{No SR} & 2.895 & \textbf{0.802} & 0.306 & \textbf{27.125} &  0.547 \\
\hline
x265  &  RVRT~\cite{liang2022recurrent} & \textbf{2.768}  & 0.751 & 0.281 & 27.021 &  \textbf{0.506} \\
x265  &  \textit{No SR} & 2.011 & \textbf{0.82}5 & \textbf{0.221} & \textbf{27.152} &  0.496 \\
x265  &  COMISR~\cite{li2021comisr} & 1.831 & 0.726 & 0.285 & 26.967 &  0.506 \\
\hline
vvenc  &  RVRT~\cite{liang2022recurrent} & \textbf{2.421}  & 0.755 & \textbf{0.295} & \textbf{27.425} &  0.501 \\
vvenc  &  RBPN~\cite{haris2019recurrent} & 1.393 & 0.752 & 0.331 & 27.424 &  \textbf{0.509} \\
vvenc  &  \textit{No SR} & 0.944 & \textbf{0.835} & 0.325  & 26.467 &  0.500 \\
\hline
uavs3e  &  \textit{No SR} & \textbf{2.313} & \textbf{0.833} &\textbf{ 0.270} & \textbf{27.174} & \textbf{ 0.505} \\
uavs3e  &  RVRT~\cite{liang2022recurrent} & 1.682 & 0.761 & 0.291 & 27.040 &  0.503 \\
uavs3e  &  RBPN~\cite{haris2019recurrent} &  1.404  & 0.737 & 0.294 & 26.998 &  0.499 \\
\hline
aomenc  &  \textit{No SR} & \textbf{2.884} &\textbf{ 0.856} & \textbf{0.235} &\textbf{ 27.238} & \textbf{0.509} \\
aomenc  &  RVRT~\cite{liang2022recurrent} & 1.752 & 0.755 & 0.283 & 27.046 &  0.501 \\
aomenc  &  RBPN~\cite{haris2019recurrent} & 1.599 & 0.730 & 0.288 & 27.018 &  0.499 \\
\hline
  \end{tabular}}
  \end{center}
  \caption{Comparison of SR+codec pairs by subjective score and objective metrics for “Restaurant” sequence.\protect\footnotemark[1]{} The best result appears in \textbf{bold}.}
  \label{tab:benchmark_pairs_rank}
\end{table}

\footnotetext[1]{Full results are available on the benchmark page: \href{https://videoprocessing.ai/benchmarks/super-resolution-for-video-compression.html}{https://videoprocessing.ai/benchmarks/super-resolution-for-video-compression.html}}

We split our comparison into five sections by codec, using only the 10 best SR models as determined by the LPIPS value. Also, from each group of models with similar architectures, like SOF-VSR-BI and SOF-VSR-BD, we selected only one model showing the best LPIPS values. 

During the experiment we showed each participant a pair of videos from two random SR models and asked them to choose the video that looks more realistic and has fewer compression artifacts (“indistinguishable” was also an option). Every video pair was viewed by 15 participants. Each participant compared 25 pairs total. 

Among the 25 questions were three verification ones, which had obvious predefined answers. We excluded the results from any participant who failed to correctly answer one or more of the verification questions. A total of 5662 people participated in our subjective evaluation. We excluded the results from 265 of them because they failed to correctly answer verification questions. Our calculation of the final subjective scores, using the Bradley-Terry model~\cite{bradley1952rank}, employed the remaining 120,316 responses.

\section{Evaluation Results}

In this section we present the results of our Super-Resolution for Video Compression benchmark. We discuss two different ways of comparing SR models in our benchmark: by evaluating the quality of the results and by the bitrate reduction compared to the baseline codec. We also analyze existing video quality metrics based on subjective evaluation.



\begin{table}[t]
\begin{center}
\begin{tabular}{l|ccccc}
\hline
\backslashbox{SR}{codec} & x264 & x265 & aomenc & vvenc & uavs3e\\
 \hline
\textit{No SR} & 1.000 & 1.000 & \textbf{1.000} & \underline{1.000} & \emph{1.000}\\
RealSR\cite{ji2020real} & \textbf{0.196} & \underline{0.502} & \emph{1.513} & 4.470 & \textbf{0.639} \\
RVRT\cite{liang2022recurrent} & \underline{0.271} & 0.724 & 2.806 & \emph{1.665} & 1.750\\
SwinIR\cite{liang2021swinir} & \emph{0.304} & \textbf{0.346} & \underline{1.505} & 2.502 & \underline{0.640}\\
Real-ESRGAN\cite{wang2021realesrgan}  & 0.335 & \emph{0.640} & 2.411 & 4.368 & 2.430\\
COMISR\cite{li2021comisr} & 0.367 & 0.741 & 2.799 & \textbf{0.701} & 1.603\\
VRT\cite{liang2022vrt} & 1.245 & 3.175 & 4.157 & 4.185 & 3.663\\
BasicVSR++\cite{chan2022basicvsr++} & 1.971 & 3.390 & 4.199 & 4.185 & 4.250\\
RBPN\cite{haris2019recurrent} & 1.979 & 3.434 & 4.226 & 4.246 & 4.470\\
\hline
\end{tabular}
\end{center}
\caption{Average BSQ-rate~\cite{zvezdakova2020bsq} over subjective scores for each SR+codec pair. Lower BSQ-rate is better. The best result appears in \textbf{bold}, the second best is \underline{underlined}, and the third best is in \textit{italics}.}
\label{tab:benchmark_subjective_results}
\end{table}

\subsection{Comparison by resulting video quality}
\label{sec:quality_assessment_comparison}

Table \ref{tab:benchmark_pairs_rank} lists two best SR methods for each codec and \textit{“No SR”} method, which applies the video codec to the source video without downscaling or super-resolving. The best methods are chosen based on subjective scores on the \textit{“Restaurant”} sequence compressed at an approximate bitrate of 2 Mbps.

We can see that RBPN and RVRT appear to be the best methods for various advanced codecs, mainly due to their video restoration capabilities. However, this is not the case for x264 codec, where SwinIR and Real-ESRGAN come on top. Videos compressed with x264 at lower bitrates have many compression artifacts that make it hard to faithfully restore the original sequence. Thus, generative SR models such as Real-ESRGAN perform subjectively better by generating realistic frames.

 We also see that \textit{“No SR”} exhibits the best results for the aomenc codec. This is likely because at low bitrates, AV1 codecs use a special mode that encodes frames at a low resolution and applies an upsampler when decoding~\cite{joshi2019loop}, making the additional use of SR redundant.

\subsection{Comparison by bitrate reduction}
The results of BSQ-rate calculation over subjective scores of each SR+codec pair appears in Table \ref{tab:benchmark_subjective_results}. As the table shows, the best SR model differs by codec, proving that no single SR model can handle distortion from all compression standards with equal effectiveness.
Although RealSR shows significantly less accurate results at high bitrates than other SR models, it gives more visually pleasing results at low bitrates, which are subjectively comparable to \textit{“No SR”} results at high bitrates. We can see that \textit{“No SR”} shows better results with aomenc codec for the same reason we discussed in the previous subsection.

\begin{figure}
\begin{minipage}[c]{0.45\linewidth}
\vspace{0pt}
  \centering
  \includegraphics[width=.85\linewidth]{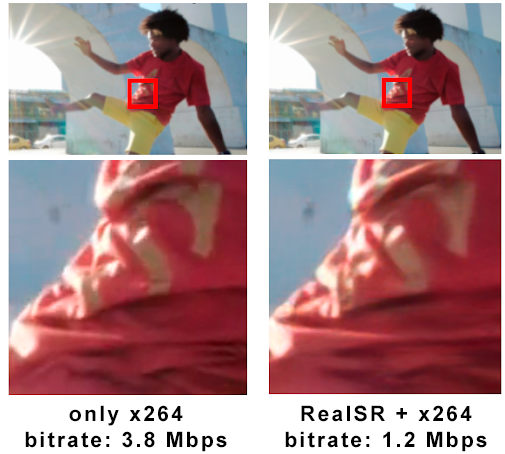}
  \caption{RealSR applied to video compressed with x264 codec at 1.2 Mbps can achieve the same visual quality as plain x264 codec at 3.8 Mbps.}
  \label{fig:bitrate_reduction}
\end{minipage}%
\hspace{0.05\linewidth}
\begin{minipage}[c]{0.45\linewidth}
\vspace{0pt}
  \centering
  \begin{tabular}{l|cc}
\hline
    Metric & PLCC & SRCC \\
\hline
    MS-SSIM~\cite{wang2003multiscale} & 0.146 & 0.151\\
    PSNR & 0.187 & 0.285 \\
    VMAF~\cite{VMAF_repo} & 0.344 & 0.448 \\
    LPIPS~\cite{zhang2018unreasonable} & 0.414 & 0.431 \\
    ERQA~\cite{kirillova2021erqa} & 0.582 & 0.624 \\
    MDTVSFA~\cite{li2021unified} & 0.634 & 0.644 \\
\hline
    ERQA$\times$MDTVSFA & \textbf{0.770} & \textbf{0.801} \\
\hline
  \end{tabular}
\vspace{2.5pt}
\captionof{table}{Mean Pearson (PLCC) and Spearman (SRCC) rank correlation coefficients between metrics and subjective-comparison results.}
\label{tab:metric_correlation}
\end{minipage}
\end{figure}

\subsection{Video quality metrics assessment}
\label{sec:metric_correlation_section}

Judging by the results in Section \ref{sec:quality_assessment_comparison},  some video quality metrics cannot reproduce the result of the subjective evaluation obtained by crowd-sourcing. To analyze the objective metrics, we calculated them on the crops that were used for the crowd-sourced evaluation. Pearson (PLCC) and Spearman (SRCC) correlations of objective metrics with subjective scores are presented in Table \ref{tab:metric_correlation}.

We analyzed the performance of the best metrics and noticed that ERQA shows a high correlation on the video crops where MDTVSFA has a low correlation, and vice versa. We believe that these two metrics take into account different features of the input video and can complement each other. Since ERQA values range from 0 to 1, ERQA can be used as a correction value for MDTVSFA. We attempted to combine the ERQA and MDTVSFA metrics by multiplying their results. We called this method ERQA$\times$MDTVSFA. This approach yielded a significant increase in correlation, as shown in Table \ref{tab:metric_correlation}.


\section{Conclusion}
In this paper we proposed a new benchmark for super-resolution (SR) compression restoration. Our work assessed 19 SR models applied to five video codecs using both objective-quality metrics and subjective evaluation.

Our research shows that SR models, such as RealSR~\cite{ji2020real} and RVRT~\cite{liang2022recurrent}, can serve in downscale-based codecs on the decoder side to enhance the subjectively perceived quality of videos with low bitrates. RVRT can improve the quality of videos compressed by x265 and vvenc codecs. RealSR can be used with x264 to reduce the video bitrate by more than 65\% without significant quality loss, as shown in Figure~\ref{fig:bitrate_reduction}. 

We find that existing video-quality metrics correlate poorly with subjective scores and are therefore unsuitable for assessing the results of downscale-based video coding. MDTVSFA and ERQA show better results, and their combination ERQA$\times$MDTVSFA achieves Spearman's rank correlation of 0.801 on our dataset.

Our benchmark is publicly available at \href{https://videoprocessing.ai/benchmarks/super-resolution-for-video-compression.html}{videoprocessing.ai/benchmarks/super-resolution-for-video-compression.html}. We welcome SR researchers to contribute to it by submitting SR models.



\bibliography{egbib}
\end{document}